\documentclass{article}
\usepackage{graphicx}
\input amssym.def
\input amssym.tex
\def\p{\partial}
\def\lb{\label}
\def\ben{\begin{eqnarray}}
\def\een{\end{eqnarray}}
\def\be{\begin{eqnarray}}
\def\ee{\end{eqnarray}}
\def\half{{1 \over 2}}
\def\bea{\begin{eqnarray}}
\def\eea{\end{eqnarray}}

\newcount\hour \newcount\minute
\hour=\time  \divide \hour by 60 \minute=\time \loop \ifnum
\minute > 59 \advance \minute by -60 \repeat
\def\nowtwelve{\ifnum \hour<13 \number\hour:
                      \ifnum \minute<10 0\fi
                      \number\minute
                      \ifnum \hour<12 \ A.M.\else \ P.M.\fi
         \else \advance \hour by -12 \number\hour:
                      \ifnum \minute<10 0\fi
                      \number\minute \ P.M.\fi}
\def\nowtwentyfour{\ifnum \hour<10 0\fi
                \number\hour:
                \ifnum \minute<10 0\fi
                \number\minute}

\begin{document}

\begin{flushright}
DAMTP-2007-22 \ \ \ \
\end{flushright}

\vspace{10pt}
\begin{center}

{\LARGE The Geometry of Small Causal Diamonds}

\vspace{20pt}

{\large G. W. Gibbons$^{1,3}$ and  S. N. Solodukhin$^{2,3}$}

\vspace{20pt}

{$^1$ D.A.M.T.P., Cambridge University, Wilberforce Road,\\
Cambridge CB3 0WA, U.K.}\footnote{Permanent address}\\

\vspace{5pt}

{$^2$ School of Engineering and Science,  International University
Bremen, \\
P.O. Box 750561, Bremen 28759,
Germany$^*$}\\

\vspace{5pt}

{$^3$ Institut des Hautes \'Etudes Scientifiques, \\
35, route de
Chartres, 91440 Bures-sur-Yvette, France}

\end{center}
{ \vspace{10pt}

\vspace{40pt}

\begin{abstract}
The geometry of causal diamonds or Alexandrov open sets whose
initial and final events $p$ and $q$ respectively have a
proper-time separation $\tau$  small compared with the curvature
scale is a universal. The  corrections from flat space are  given
as a power series in $\tau$ whose coefficients   involve  the
curvature at the centre of the diamond. We  give formulae for the
total  4-volume $V$ of the diamond, the area $A$ of the
intersection the future light cone of $p$ with  the past light
cone of $q$ and the 3-volume of the hyper-surface of largest
3-volume bounded by this intersection valid to ${\cal O } ( \tau
^4) $. The formula for the 4-volume agrees with a previous  result
of Myrheim. Remarkably, the iso-perimetric ratio ${3V_3 \over 4
\pi } / ({ A \over 4 \pi } ) ^{3 \over 2}  $  depends only on the
energy density at the centre and is bigger than unity if the
energy density is positive. These results are also shown to hold
in all spacetime dimensions. Formulae are also given, valid to
next non-trivial order, for causal domains in two spacetime
dimensions.

We suggest a number of applications, for instance, the directional
dependence of the volume  allows one to regard the volumes of
causal diamonds as an  observable providing   a measurement of the
Ricci tensor.

\end{abstract}

\newpage

\section{Introduction}
Causal Diamonds, or Alexandrov open sets, play an increasingly
important role in quantum gravity, for example in the approach via
casual sets \cite{Sorkin},  in  discussions of \lq holography\rq ,
and also of the probability of various observations in eternal
inflation models (see \cite{Bousso} for a recent example and
references to earlier work). Curiously, however, not very much  of
a quantitative  nature appears to be known about them. The purpose
of this note is to embark on a remedy of that situation at least
in the case that  the diamond is small compared with the curvature
scale of the  ambient spacetime. In fact the causal structure and
volume measure are sufficient to fix the spacetime topology,
differential structure, and  metric completely
\cite{HawkingKingMcCarthy,Malament}. This information can be
encoded in a knowledge of the set of small  causal diamonds and
their {\sl volumes}.  In particular, it allows one to extract not
only the metric, but the Ricci tensor as well allowing one to
formulate the vacuum Einstein equations in a simple way
\cite{Myrheim}. The  volume of causal diamonds plays an important
role  in studies  of eternal inflation where one introduces a
single connected {\it Meta-Universe} \footnote{sometimes called a
{\it Multiverse}}   and takes the probability for the occurrence
of a `pocket'  in the Meta-Universe to be the volume of an
appropriate causal diamond.

\section{Causal Diamonds}

A {\it causal diamond}  or {\it Alexandrov open set} is a subset
of a Lorentzian spacetime $\{M, g\}$ of the form \ben I^+(p) \cap
I^-(q)\,, \een where $I^+,I^- $ denotes chronological future or
past respectively. The causal diamond   depends only on the
conformal class of the Lorentzian metric $g$ but the 4-volume for
example \ben V(p,q) = {\rm Vol}( I^+(p) \cap I^-(q     )= \int _{
I^+(p) \cap I^-(q) } \sqrt{|g|} d^4\,x \,, \een depends upon the
metric itself. In Minkowski spacetime ${\Bbb E} ^{3,1}$,
\footnote{we use signature $-+++$ and MTW/HE  curvature
conventions throughout} as long as $p$ is in the chronological
past of $q$, there exists a unique straight line, i.e. time-like
geodesic, joining them, and if its proper lengths  is $\tau$,
then, as pointed out by 't Hooft \cite{'tHooft}
 \ben V(p,q) ={\pi \over 24} \tau ^4\,.
\label{'tHooft}\een

In a general curved spacetime,  if $p$ and $ q$ are sufficiently
close, there will still be a unique time-like geodesic
$\gamma(t)$, parametrised by proper time $ t \in [-{\tau \over 2},
{\tau \over 2}]$. joining them as long as  $p$ remains  in the
chronological past of $q$. 't Hooft's  formula (\ref{'tHooft})
will be approximately true but there are corrections involving the
curvature evaluated at the mid point $0=\gamma(0) $.

One way of calculating them is to use Riemann normal coordinates
${x^\mu}$ centred on $0$, in which the metric takes the form \ben
g_{\mu \nu}= \eta_{\mu \nu} -{ 1 \over 3} R_{\mu \sigma \nu \tau}
(0)  x^\sigma x^\tau +\dots \qquad . \een

This was done by Myrheim \cite{Myrheim} who obtained the result
\ben V(p,q) ={\pi \over 24} \tau ^4 \Bigl(1 + a \tau ^2 R(0) + b
R_{\mu \nu}(0)
 T^\mu T ^\nu  +\dots  \Bigr ) \,,
\label{Myrheim1}\een where $T^\mu  $ is the  time-like vector
tangent to the geodesic at the origin $\gamma$ and with magnitude
\ben g_{\mu \nu}(0) T ^\mu T ^\nu = -\tau ^2\,. \een The expansion
in $\tau$ is thus also expansion in $T^\mu$.

Thus (\ref{Myrheim1}) may also be written as

\ben V(p,q) ={\pi \over 24} \tau ^4 \Bigl(1 + a \tau ^2 R(0) + b
\tau ^2
 R_{\hat 0 \hat 0}(0) +\dots  \Bigr ) \,,
\label{Myrheim2}\een where $R_{\hat 0 \hat 0}(0)$ are the
time-time components of the Ricci tensor at the origin evaluated
in an orthonormal frame at the origin whose zero leg is aligned
with the tangent vector of the geodesic $\gamma$.

One may also express $ T ^\mu $ in terms of Synge's {\it World
Function} $\Omega(p,q) $ giving the distance squared between two
events $p,q$ and  which satisfies \ben g^{\mu \nu} \p _\mu \Omega
\p _ \nu \Omega = -4\Omega \,. \label{Synge} \een One has \ben
T^\mu = \half g^{\mu \nu } \p_\nu \Omega\,. \een If $p$ is in the
past of $q$ one may define $\tau (p,q)= \sqrt{\Omega (p,q)} $ and
(\ref{Synge}) becomes the Hamilton-Jacobi equation \ben g^{\mu
\nu} \p_\mu  \tau   \p _\nu \tau =1\,. \een

Myrheim did not give the full details of his calculation in
Riemann normal coordinates. Indeed the derivation is complicated
by the need to find the curvature induced deviation of the light
cones from their flat space positions. To our knowledge no other
derivation has been given since. Because  the coefficients $a$ and
$b$ in  Myrheim's formula are  universal, i.e. valid for {\sl any}
spacetime, one should be able to avoid the use of Riemann normal
coordinates altogether, but rather to determine the coefficients
by considering two special cases. This is what we shall now do.
Using the same technique we shall also obtain new results for the
area $A(p,q)$ of the intersection of the past and future light
cones, and for maximal three-volume $V_3(p,q)$ of any hypersurface
which it bounds.

\subsection{The Einstein Static Universe}

The metric is \ben ds^2 =- dt ^2 + d \chi ^2 + \sin ^2 \chi \bigl
(d \theta ^2 + \sin^2 \theta \phi ^2  \bigl)\,. \een We take $p$
as $(-\half \tau, 0,0,0)$ and $q$  as
 $(\half \tau, 0,0,0)$.
 Since the metric is an unwarped (i.e ultrastatic)
 product of the  unit round
three-sphere with time,
 $R_{\hat 0 \hat 0}(0)   =0 $ and $R=6$,
the value for the unit three-sphere.

The past light cone of $q$ is given by $t= {\tau \over 2} - \chi$
and the future light cone of $p$ by $t= -{\tau \over 2} + \chi$.

The volume is easily seen to be \ben V(p,q) =  V(\tau) &=&
 8 \pi \int _0 ^{\half \tau} dt \Bigl( \int _0 ^{ \half \tau - t}
\sin^2 \chi d \chi  \Bigr)\, \nonumber \\
&=&{\pi\over 2}(\cos^2{\tau\over 2}+{\tau^2\over 4}-1)\,.
 \een
The small $\tau$ expansion  of $V(\tau)$ then can be obtained to
arbitrary order,
 \ben V(\tau)= {\pi\tau ^4 \over 24} \bigl( 1
-{\tau ^2 \over 30} +\dots \bigr) \,.\een
 This fixes, in our
conventions, the constant $a$ to be \ben a= - {1 \over 180}\,.
\een and agrees up to a sign with the expression given by Myrheim
\cite{Myrheim}.

\subsection{de-Sitter spacetime}

The metric is \ben ds^2 =-dt^2 + \cosh^2 t \Bigl(  d \chi ^2 +
\sin ^2 \chi \bigl (d \theta ^2 + \sin^2 \theta \phi ^2  \bigl)
\Bigr )  \,. \een We take $p$ as $(t= -\half \tau, 0,0,0)$ and $q$
as
 $(t= \half \tau, 0,0,0)$
The Ricci scalar $R=12$ and since $R_{\mu \nu} = 3 g_{\mu \nu}$,
then   $R_{\hat 0 \hat 0}(0)   =-3$.

 To obtain the past light cone we introduce conformal time $\eta$ by
\ben d \eta = { dt \over \cosh t}\,, \een choosing the constant of
integration so that at $t=0$ , $\eta =0$, one finds that \ben
\tan( {\pi \over 4} + {\eta \over 2} ) = e^t \,. \een Further
useful relations are \ben \sinh t =\tan \eta\,,\qquad  \cosh t = {
1 \over \cos \eta } \,.\een

The metric is now
 \ben ds^2 =  { 1 \over \cos ^2  \eta} \Bigl (
-d \eta ^2 +  d \chi ^2 + \sin ^2 \chi \bigl (d \theta ^2 + \sin^2
\theta \phi ^2 \bigl) \Bigr )  \,. \een Define  $N$ by \ben \cosh
{\tau \over 2} = {1\over \cos { N \over 2}} \,, \een then  $p$ is
at $(\eta = -\half N, 0,0,0)$ and $q$ is at
 $(\eta = \half N, 0,0,0)$
The past light cone of $q$ is given by $\eta = {N \over 2} - \chi$
and the future light cone of $p$ by $\eta = -{N \over 2} + \chi$.

The volume is easily seen to be \ben V(p,q) &=&  V(\tau) =
 8 \pi \int _0 ^{\half N } { d \eta \over \cos ^4 \eta}
 \Bigl( \int _0 ^{ \half N - \eta}
\sin^2 \chi d \chi  \Bigr)\,\nonumber \\
&=&{4\over 3}\pi (\cos^2{N\over 2}-2\ln\cos{N\over 2} -1)\nonumber
\\
&=&{4\over 3}\pi ({1\over \cosh^2{\tau\over 2}}+2\ln
\cosh{\tau\over 2}-1)\,. \label{V}
 \een
This function can be expanded to any order in $\tau$. We are
however interested in the  first few terms,
 \ben V(\tau) = { \pi
\over 24} \tau ^4 \bigl ( 1 - {\tau ^ 2\over 6} +\dots \bigr )\,.
\een
 This gives the  value  \ben b = { 1 \over 30}\, \een
for the coefficient $b$,  which  agrees, up to a sign convention,
with the result of Myrheim who uses the opposite signature to us.
His metric is thus minus our metric and hence, although his Ricci
tensor is the same as ours, his Ricci scalar has the opposite
sign.

On the other hand, in both cases the volume  has an interesting
behavior for large $\tau$. One has that
 \ben V(\tau)={4\over 3}\pi \tau+O(1)
\een for de Sitter space-time and \ben V(\tau)={\pi\over 2}\tau^2
+O(1)\een for the Einstein Static Universe.
 Notice that in both cases  the leading term grows much
slower than $\tau^4$.This presumably reflects the fact that both
spacetimes satisfy the null convergence condition,
 $R_{\mu \nu}  l^\mu l^\nu \ge 0$ for all light-like vectors $l^\mu$.

\section{Area}

The area $A(p,q)$ of the intersection $\dot I^+(p) \cap \dot
I^-(q) $ of the future light cone of $p$, ${\dot I}^+(q)=\p
I^+(p)$ with the past light cone of $q$, ${\dot I}^+(q)=\p I^+(p)$
is also given by a universal formula which to next to lowest order
might be expected to depend on both  the Ricci scalar $R$ and
$R_{\hat 0\hat 0}$ , the time-time component of the Ricci tensor.

In both our examples the intersection is on the surface $t=0$ and
we have \bea A(p,q)&=& 4\pi \sin ^2 (\half N)\,.
\label{A1}
 \eea For the Einstein Static Universe one has $N=\tau$
and hence
\ben A(p,q)&=& 4 \pi \sin^2{\tau\over 2}\nonumber \\
&=& \tau ^2 \bigl (1 - {1 \over 12} \tau^2 +\dots \bigr ) \,.\een
For de-Sitter spacetime we have a relation
$\cos(N/2)=1/\cosh(\tau/2)$ and  thus
\ben A(p,q)&=&4\pi\tanh^2({\tau\over 2})\nonumber \\
&=& \pi \tau ^2 \bigl (1 - {1 \over 6 } \tau^2 +\dots \bigr )\,
\label{A2} \een These imply that in general \ben A(p,q)= {\rm
Area} ( \dot I^+(p) \cap \dot I^-(q) )=
   \pi \tau ^2 \bigl (1 - {1 \over 72 }R  \tau^2  +\dots \bigr )\,.
\een

Thus the area $A(p,q)$, unlike the 4-volume $V(p,q)$, contains no
directional information. Notice, however, that the area is given
by  different functions  (\ref{A1}) and (\ref{A2}) of $\tau$ in
these two cases. This is due to the presence of powers of the
Ricci tensor in the $\tau$ expansion of the area which show up in
the higher order terms.

\section{Three-volume}

There are infinitely many space-like hypersurfaces having the
intersection  $\dot I^+(p) \cap \dot I^-(q) $ of the future light
cone of $p$ with the past light cone of $q$ as their  boundary.
Among them, provided $p$ and $q$ are sufficiently close, there is
one   with maximal volume \footnote{we remind the reader that
spacelike hypersurfaces in Lorentzian spacetimes can have
arbitrarily small volumes and hence  the concept of a minimal
spacelike hypersurface is not well defined}. The maximal value of
this 3- volume $V_3(p,q)$ should also be given by a universal
formula.

Our two examples are time symmetric and therefore the hypersurface
of maximal volume has $t=0$. Thus \ben V_3(p,q) &=&4\pi
\int_0^{N\over 2}d\chi \sin^2\chi \nonumber
\\
&=& \pi (N-\sin N) \, .\een Using the already established
relations between $N$ and $\tau$ we have that \ben V_3(\tau)=\pi
(\tau-\sin\tau)\een in the case of the Einstein Universe and \ben
V_3(\tau)=2\pi\left(\arctan(\sinh{\tau\over 2})-{\sinh{\tau\over
2}\over \cosh^2{\tau\over 2}}\right)\een in the case of de-Sitter
space-time. Therefore  for the Einstein Static Universe, \ben
V_3(p,q) = { \pi \over 6} \tau ^3 \bigl( 1 - {1 \over 20} \tau^2
+\dots \bigr )\,, \een while for de-Sitter spacetime \ben V_3(p,q)
= { \pi \over 6} \tau ^3 \bigl( 1 - {7 \over 40} \tau^2  +\dots
\bigr )\,. \een

Therefore, in general, \bea V_3(p,q)& =& { \pi \over 6} \tau ^3
\bigl( 1 - {1  \over 120} R
\tau^2  +  { 1 \over 40} R_{\hat 0 \hat 0}  \tau ^2 + \dots \bigr )\nonumber\\
& =& { \pi \over 6} \tau ^3 \bigl( 1 - {1  \over 120 } R \tau^2  +
{ 1 \over 40} R_{\mu \nu  } T^\mu T^\nu   \dots \bigr  )\,. \eea

\section{Energy Conditions}

We have shown that \bea V(p,q) &=& {\pi \over 24 }  \tau ^4 \Bigl
( 1 + { 1 \over 180} \bigl( R g_{\mu \nu}   +
6  R_{\mu \nu } \bigr )  T^\mu T^\nu   +\dots \Bigr )\nonumber \\
&=& {\pi \over 24 }  \tau ^4 \Bigl ( 1  + {4 \pi G \over 45 }
\bigl( 3 T_{\mu \nu} -2 T g_{\mu \nu } \bigr )  T^\mu T^\nu +\dots
\Bigr )\,. \eea
 In the last line we have used the Einstein
equations \ben R_{\mu \nu} = 8 \pi G \bigl( T_{\mu \nu }- \half T
g_{\mu \nu}  \bigr )\,, \een where $T_{\mu \nu}$ is the
energy-momentum tensor.

As an illustration rather than a check, one may substitute the
energy momentum  tensor of an inflating universe \ben T_{\mu \nu}
= -V(\phi) g_{\mu \nu} \,, \een where $V(\phi)$ is the potential
energy of an inflation field $\phi$ to get \ben V(p,q) = {\pi
\over 24 }
 \tau ^4 \Bigl(  1- {4 \pi G  V(\phi) \over 9} \tau ^2 +\dots  \Bigr )
\een which of course agrees with the de-Sitter results if one uses
the relations \ben \Lambda = 8 \pi G V(\phi) =3 \,. \een

In the presence of a perfect fluid whose velocity is aligned with
the time-like tangent vector of the geodesic $\gamma$ , one finds

\ben V(p,q) = {\pi \over 24 } \tau ^4 \Bigl(  1 +  {4 \pi G \over
45} (\rho + 6 P )
 \tau ^2 +\dots  \Bigr )\,,
\een where $\rho$ is the energy density and $P$ the pressure of
the fluid. If both are positive, the causal diamond has a larger
volume, for fixed proper time duration $\tau$ than it would have
in flat spacetime. By contrast for large negative pressures, as
during inflation, the volume is smaller than it would be in flat
spacetime.

\section{Isometric Inequalities}

In flat Euclidean space ${\Bbb E} ^3$ the 3-ball maximises volume
enclosed for fixed surface area. It is thus of interest to examine
the iso-perimetric ratio \ben {3V \over 4 \pi } / \bigl ({ A \over
4 \pi }\bigr ) ^{3 \over 2} \,. \een One has \bea {3V_3(p,q) \over
4 \pi } / \bigl ({ A(p,q)
 \over 4 \pi }\bigr ) ^{3 \over 2} & =&  \Bigl ( 1+{ 1\over 80} R \tau ^2
+ { 1 \over 40} R_{\hat 0\hat 0} \tau^2 +\dots \Bigl ) \nonumber \\
& =&  \Bigl ( 1+ {1 \over 40} \bigl( R_{\mu \nu} -\half R g_{\mu
\nu}
\bigr ) T^\mu T^\nu   +\dots \Bigl )\nonumber \\
& =&\Bigl ( 1+ { \pi G \over 5 }   T_{\mu \nu}   T^\mu T^\nu
+\dots \Bigl )\,, \eea where in the last line we have used the
Einstein equations. Remarkably, the iso-perimetric ratio involves
just the energy density at the origin $O$ of the diamond and
exceeds unity if the Weak Positive Energy Condition holds.

Relation of a similar type which involves the 4-volume $V(p,q)$ is
\be {24\over \pi} V(p,q)/({6 V_3(p,q)\over \pi})^{4\over
3}=(1+{1\over 180}R \tau^2+\dots)\, . \lb{ratio}\ee As we see the
directional part cancels in (\ref{ratio}) so that the
iso-perimetric ratio (\ref{ratio}), at least at the given order in
$\tau$,  is direction independent. Whether this ratio exceeds
unity depends only on the sign of the Ricci scalar, similarly to
the iso-perimetric ratio for Euclidean manifolds.

\section{Measuring the Ricci and the Riemann tensors}

The formula for $V(p,q)$ should be compared with that for the
volume $V(r) $ of a 4-ball of radius $r$
 in a Riemannian  manifold
\ben V(r)  = {\pi ^2 \over 2} r^4  \Bigl( 1- {1 \over 36 } R r^2
+\dots \Bigr ) \label{Vball} \,. \een There is an important
difference in that the formula for the volume of a causal diamond
is directional in character since it involves not only the Ricci
scalar $R$, but the Ricci tensor $R_{\mu \nu}$. By varying the
points $p$ and $q$ and hence varying $T ^\mu $ one could, if one
could measure $V(p,q)$, determine the whole Ricci tensor for
vectors within the light cone. Since it is a bilinear function on
tangent vectors, its value for space-like vectors is then given by
continuity and, assuming continuity, it is thus uniquely
determined.

Indeed, given a causal structure and some measure of volume and
proper time, one could use the formula to {\sl define} the Ricci
tensor. This might be useful in approaches to quantum gravity
based on Causal Sets or Directed Graphs.

The same remarks apply to the three-volume $V_3(p,q)$. However, as
remarked earlier,  the area $A(p,q)$ would only allow one to
measure the Ricci scalar $R$.

The formula (\ref{Vball}) can be generalised. Let's take an
$n$-dimensional cycle $\Sigma_n$ and  consider a tube $T(\Sigma_n,
r)$ of radius $r$ in the direction orthogonal to $\Sigma$. The
volume of this tube has a small $r$ expansion, \ben
V(T)=V(B_{4-n}(r))\int_{\Sigma_n}\left(1+c_n
\sum_{i,j=1}^{4-n}R_{ijij}r^2+\dots\right)\, , \label{VT} \een
where $V(B_{4-n}(r))$ is volume of $(4-n)$-dimensional ball of
radius $r$ in flat space and
$R_{ijij}=R_{\alpha\mu\beta\nu}n^\alpha_i n^\beta_i n^\mu_j
n^\nu_j$, where $n_i^\mu$, $i=1..4-n$ is a set of orthonormal
vectors orthogonal to $\Sigma_n$. When $n=0$ ($\Sigma$ is just a
point) one has that $\sum_{ij}R_{ijij}=R$ and (\ref{VT}) becomes
(\ref{Vball}).

In the Lorentzian signature it is straightforward to introduce an
analogous construction of a `causal tube' for  cycle $\Sigma_n$
lying entirely in a space-like hypersurface. The causal tube then
can be defined as the union of causal diamonds of `size' $\tau$
centered at every point of $\Sigma$. The causal structure of the
tube is determined by the field of time-like vectors $T^\mu$
defined everywhere on $\Sigma_n$. To leading order in $\tau$ the
volume of this causal tube is a product of  volume
$V_{(4-n)}(\tau)$ of ($4-n)$-dimensional causal diamond and the
volume of $\Sigma_n$,
$$
V(T)=V_{(4-n)}(\tau){\rm Vol}(\Sigma_n)\, .
$$
The higher order in $\tau$ corrections then involve components of
the Riemann tensor in the directions orthogonal to $\Sigma_n$
similar to (\ref{VT}),
 \ben
V(T)=V_{(4-n)}(\tau)\int_{\Sigma_n}\left(1+ \sum_{i,j=0}^{3-n}(a_n
R_{ijij}+b_nR_{i0i0}) \tau^2+\dots\right)\, . \label{VT1} \een The
set of vectors $n_{i}, \ i=0..(3-n)$ orthogonal to $\Sigma_n$
includes the time-like vector $ n^\mu_0=\tau^{-1}T^\mu$. The exact
values of coefficients $a_n$ and $b_n$ are to be determined.

Obviously, using these tubes one could measure components of the
Riemann tensor including those in the space-like directions.

In order to check the formula (\ref{VT}) let us consider the
Euclidean Schwarzschild metric and take horizon sphere as the
cycle, $n=2$ in this case. The Schwarzschild metric can be brought
to the form \ben ds^2=d\rho^2+g(\rho)d\phi^2+r^2(\rho)d\omega^2\,
, \een where coordinate $\phi$ has period $2\pi$ and functions
$g(\rho)$ and $r(\rho)$ are given by expansion \be
g(\rho)&=&\rho^2-{1\over 3
a^2}\rho^4+O(\rho^6)\, , \nonumber \\
r(\rho)&=&a+{\rho^2\over 4a} +O(\rho^4) \, ,\ee where $a$ is
radius of the horizon.

The volume of tube $T$ of radius $r$ (in the direction orthogonal
to horizon $\Sigma$) is given by expression \be V(T)&=&4\pi a^2
2\pi \int_0^rd\rho \sqrt{g(\rho)}r^2(\rho) \nonumber \\
&=&\pi r^2 {\rm Area} (\Sigma)(1+{r^2\over 6a^2}+\dots)\, . \ee
Taking into account that for the Schwarzschild metric one has
${1\over a^2}={1\over 2} \sum_{i,j=1}^2 R_{ijij}$, where the
curvature components are calculated at the horizon, we find that
\be V(T)=\pi r^2 \int_\Sigma \left(1+{1\over
12}\sum_{i,j}R_{ijij}r^2+\dots\right)\, , \label{VS} \ee in
agreement with general formula (\ref{VT}). Combining (\ref{Vball})
and (\ref{VS}) it seems that value of coefficient $c_n={2n-1\over
36}$ fits nicely both cases.

\section{Higher Order Results}

In this section we obtain the formula for the volume, or area, of
a causal diamond in a two-dimensional spacetime which are valid to
order $\tau^6$. We start with a two-dimensional metric in general
conformally flat form \be ds^2=a^2(x,\eta)(-d\eta^2+dx^2)\, .
\lb{m1}\ee The volume of causal diamond in this metric takes the
form \be V(\tau)=\int_0^{N/2}d\eta\int_{\eta-{N\over 2}}^{{N\over
2}-\eta}dx [a^2(x,\eta)+a^2(x,-\eta)]\, , \label{V2}\ee where $N$
should be related to proper time $\tau$ measured along the
time-like geodesic.

We consider several simple particular cases.

\bigskip

{\bf Case 1.} The function $a(x,\eta)$ is function of only time,
$a(\eta)$. Then we can change the time variable to  $t=\int
d\eta\, a(\eta)$ that measures the proper time along the time-like
geodesic so that
$$
N=\int_{-\tau/2}^{\tau/2}{dt\over a(t)}\, .
$$
The volume (\ref{V2}) is symmetric function of $\tau$, i.e.
$V(-\tau)=V(\tau)$, so that only even powers of $\tau$ appear in
the expansion of $V(\tau)$ in powers of $\tau$. The integral over
$x$  in (\ref{V2}) may be  evaluated to give \be
V(N)=\int_0^{N/2}d\eta (N-2\eta)[a^2(\eta)+a^2(-\eta)] \, .\ee The
$\tau$-derivative of volume can be expressed in terms of functions
of $t$ only, not involving coordinate $\eta$, \be
\partial_\tau V(\tau)={1\over 2}[{1\over a({\tau\over 2})}+{1\over a(-{\tau\over
2})}] \int_{-{\tau\over 2}}^{\tau\over 2}dt \,a(t)\, .
\label{V22}\ee Since only even powers of $\tau$ will appear in the
expansion of the volume anyway, let us consider $a(t)$ to be even
function of $t$ given by a small $t$ expansion
$$
a(t)=1+a_1t^2+a_2t^4+\dots
$$
Substituting this expansion into equation (\ref{V22}) we find that
\be
\partial_\tau V=2({\tau\over 2})-{4\over 3}a_1({\tau\over
2})^3+[{4\over 3}a_1^2-{8\over 5}a_2]({\tau\over 2})^5+\dots \ee
The scalar curvature of metric (\ref{m1}) has the following
expansion \be R(t)=4a_1+(24a_2-4a_1^2)t^2+\dots \ee The terms in
the expansion of the volume then can be expressed in terms of
curvature and its derivatives evaluated in the center of the
diamond, \be V(\tau)=2({\tau\over 2})^2-{1\over 6}R\, ({\tau\over
2})^4+{1\over 45}(R^2-{1\over 2}R''_{\eta\eta})({\tau\over
2})^6+\dots \, ,\ee where we have used the fact that in the center
of the diamond $R''_{tt}=R''_{\eta\eta}$.

\bigskip

{\bf Case 2.} Function $a(\eta,x)$ depends only on coordinate $x$
and does not depend on time $\eta$. Normalising $a(x=0)=1$ in the
center of the diamond, we have that $N=\tau$ and  we find
 \be
 V(\tau)=2\int_0^{\tau/2}d\eta\int_{\eta-\tau/2}^{-\eta+\tau/2} dx
 \ a^2(x)
 \ee
 for the volume.
Representing function $a(x)$ in the vicinity of the center of the
diamond in terms of expansion
$$
a(x)=1+b_1x^2+b_2x^4+\dots
$$
we find the following  \be V(\tau)=2({\tau\over 2})^2+{2b_1\over
3}({\tau\over 2})^4+{2\over 15}(2b_2+b^2_1)({\tau\over 2})^6+\dots
\lb{exp}\ee expansion for the volume. The scalar curvature this
time has expansion \be R(x)=-4b_1+(-24b_2+20b_1^2)x^2+\dots \ee so
that the coefficients in expansion (\ref{exp}) can be expressed in
terms of values of the curvature and its derivatives in the center
of the diamond, \be V(\tau)=2({\tau\over 2})^2-{1\over 6}R\,
({\tau\over 2})^4+{1\over 45}(R^2-{1\over 4}R''_{xx})({\tau\over
2})^6+\dots \ee

In principle, we can not exclude the appearance of the term
$R''_{\eta x}$ in the $\tau^6$ term. The presence of this term can
not be detected from the above two cases. So that we need to
consider one more case.

\bigskip

{\bf Case 3.} Choose function $a(x,\eta)$ in the form \be
a(x,\eta)=1+c x^3\eta \, .\ee Time-like curve $x=0$ is still a
geodesic in this case so that $\eta$ measures the proper time
along this geodesic, and we have that $N=\tau$. The Ricci scalar
has expansion
$$ R=12c\, \eta \, x+\dots
$$
in this case so that $R''_{\eta x}=12c$.

We find that volume \be
V(\tau)=2\int_0^{\tau/2}d\eta\int_{\eta-\tau/2}^{-\eta+\tau/2}dx
a^2(x,\eta)=2({\tau\over 2})^2+{c^2\over 630}({\tau\over
2})^{10}+\dots \ee has expansion with vanishing term $\tau^6$.
This indicates that term $R''_{\eta x}$ does not appear in this
order.

\bigskip

Combining all these cases  the expansion of the volume can be
presented in the following covariant form \be V(\tau)=
2({\tau\over 2})^2-{1\over 6}R\, ({\tau\over 2})^4 +{1\over
180}[-g_{\alpha\beta}R^2+{1\over 4}(g_{\alpha\beta}\nabla^2 R-3
\nabla_\alpha\nabla_\beta R)\, ]T^\alpha T^\beta ({\tau\over
2})^4\, , \lb{1}\ee where we  have used that
$\nabla^2R=-R''_{\eta\eta}+R''_{xx}$ in the center of the diamond.

\bigskip

In the two-dimensional case it is straightforward to calculate the
volume (length) of the maximal space-like hypersurface that bounds
the intersection of the future light cone of $p$ with the past
light cone of $q$. This volume, $V_1(p,q)$ is the two-dimensional
analog of $V_3(p,q)$ considered above. We omit the details of this
calculation that follows same lines as the above calculation of
the volume of the causal domain. Here is the final result valid up
$\tau^5$ order, 
\be V_1(p,q)=\tau-{1\over 6}R({\tau\over
2})^3+{1\over 192}[-g_{\alpha\beta}R^2+{2\over
5}(g_{\alpha\beta}\nabla^2 R- 2\nabla_\alpha\nabla_\beta R)\,
]T^\alpha T^\beta ({\tau\over 2})^3\, .\label{2}\ee The formulae
(\ref{1}) and (\ref{2}) give us an idea on how complicated can be
the next to non-trivial order terms in higher dimensions where the
number of possible combinations  built out of curvature and its
derivatives is larger than in two-dimensions. This also allows us
to guess the possible structure of such terms in higher
dimensions. The direct calculation of the higher order terms in
four spacetime dimensions, however, requires more efforts and will
be reported elsewhere.

\bigskip

\noindent {\bf Note (added 12/08/2015):}
In flat $d=2$ spacetime one has that 
$$
V(\tau)=\frac{1}{2}V_1(\tau)\tau\ , \ \ \ \ \ \ \ \ \ \ \ \ \ \  \ \ \ \ \ \ \ \ \ \ \  \ \ \ \ \ \ \ \ \ \ \ \ \ \ \ \ \  \ \ (N1)
$$
where $V(\tau)$ is volume of a diamond of duration $\tau$ and $V_1$ is the ``volume'' of maximal hypersurface inside the diamond.
In a curved spacetime  one finds that (using eqs. (65) and (66) of the present  paper)
$$
\frac{V(\tau)}{\frac{1}{2}V_1(\tau)\tau}=1+\frac{1}{1440}(R^2+2\nabla^2 R)(\frac{\tau}{2})^4+\dots\ . \ \ \ \ \ \ \ \ (N2)
$$
It is interesting to note that  1) the term linear in curvature $R$ cancels in (N2);
2) the dependence on direction of vector $T^\alpha$ cancels in the ratio (N2).
This happens only in $d=2$ case since for $d>2$ both $R$ and $R_{00}$ terms are still present in the respective ratio $V/(\frac{1}{d}V_{d-1}\tau )$ 
as one can see from (73) and (77) of the next section.

\section{Higher Dimensions}

In this section we give some results valid in arbitrary dimension
$d$. We follow the strategy outlined in the first part of the
paper: consider two particular cases of the d-dimensional Einstein
Static Universe and d-dimensional de-Sitter spacetime. These two
cases help us to fix the coefficients in the small $\tau$
expansion for the volume of the causal diamond just in the same
way as it was demonstrated earlier in this paper for dimension
$d=4$.

The metric of d-dimensional Einstein Static Universe is \be
ds^2=-dt^2+d\chi^2+\sin^2\chi d\omega^2_{S_{d-2}} \lb{M1}\ee and
the metric of d-dimensional de-Sitter spacetime is \be
ds^2=-dt^2+\cosh^2t(d\chi^2+\sin^2\chi d\omega^2_{S_{d-2}})\,
,\lb{M2} \ee where $d\omega^2_{S_{d-2}}$ is standard metric on
d-dimensional sphere of unite radius.

\bigskip

\noindent{\it The volume of the causal diamond.} For metric
(\ref{M1}) the volume of the causal diamond is given by

\be V_{\rm
Einst}=2Vol(S_{d-2})\int_0^{\tau/2}dt(\int_0^{\tau/2-t}\sin^{d-2}\chi
d\chi )\, , \ee where  $Vol(S_{d-2})$ is volume of d-sphere of
unite radius. The volume in the case of de Sitter spacetime is
given by expression

\be V_{\rm dS}=2Vol(S_{d-2})\int_0^{N/2}{d\eta\over
\cos^d\eta}(\int_0^{N/2-\eta}\sin^{d-2}\chi d \chi)\, , \ee the
time-like coordinate $\eta$ is defined by relation
$\cos\eta=1/\cosh t$.

Expanding in both cases the volume in powers of $\tau$ and taking
into account that $N=\tau(1-{1\over 24}\tau^2+\dots)$ in the case
of de Sitter spacetime we find that \be V_{\rm Einst}&=&V_{\rm
flat}(\tau)\left(1-{d(d-1)(d-2)\over
24(d+2)(d+1)}\tau^2+\dots\right)\, ,\lb{V1} \\
 V_{\rm dS}&=&V_{\rm flat}(\tau)\left(1-{d(d-1)\over
 12(d+2)}\tau^2+\dots\right)\, ,
 \label{v2}
 \ee
where
$$
V_{\rm flat}(\tau)=Vol(S_{d-2}){2\over d(d-1)}({\tau\over
2})^{d}
$$
is volume of the causal diamond in d-dimensional flat spacetime.

We have to take into account that $R=(d-1)(d-2)$ and $R_{00}=0$ in
the case of metric (\ref{M1}) and $R=d(d-1)$ and $R_{00}=-(d-1)$
in the case of metric (\ref{M2}). Combined with this the equations
(\ref{V1}) and (\ref{v2}) are presented in the form \be V=V_{\rm
flat}(\tau)\left(1-{d\over 24(d+1)(d+2)}R\tau^2+{d\over
24(d+1)}R_{\hat{0}\hat{0}}\tau^2+\dots\right) \, .\ee This gives
us a formula for the volume of the causal diamond valid for any
dimension $d$.

\bigskip

\noindent{\it Volume of the maximal spacelike hypersurface.} The
volume of the spacelike hypersurface  (at $t=0$) for both
spacetimes is \be V_{d-1}(\tau)=Vol(S_{d-2})\int_0^{N\over 2}d\chi
\sin^{d-2}\chi \ee where $N=\tau$ for the Einstein Static Universe
and $N=\tau(1-{1\over 24}\tau^2+\dots)$ for de Sitter spacetime.
The calculation is straightforward and here is the result

\be V_{\rm Einst}&=&V^{(d-1)}_{\rm
flat}(\tau)\left(1-{(d-2)(d-1)\over
24(d+1)}\tau^2+..\right)\, ,\\
V_{\rm dS}&=&V^{(d-1)}_{\rm flat}(\tau)\left(1-{(d-1)(2d-1)\over
24(d+1)}\tau^2+..\right)\, , \ee where $V^{(d-1)}_{\rm
flat}(\tau)={1\over d-1}Vol(S_{d-2})({\tau\over 2})^{d-1}$.

These two expressions help us to determine the coefficients in the
expansion of the $(d-1)$-volume

\be V^{(d-1)}(\tau)=V^{(d-1)}_{\rm flat}(\tau)\left(1-{1\over
24(d+1)}R\tau^2+{(d-1)\over 24(d+1)}R_{00}\tau^2+\dots \right)\, .
\ee

\bigskip

\noindent{\it The Area.} The area $A(p,q)$ of the intersection
$\dot I^+(p) \cap \dot I^-(q) $ is given by expression \be
A=Vol(S_{d-2})\sin^{d-2}{N\over 2} \, ,\ee where $N=\tau$ for the
Einstein Static Universe and $N=\tau(1-{1\over 24 }\tau^2+\dots)$
for de Sitter spacetime. Thus, we get the expansion 
\be A_{\rm
Einst}(\tau)&=&A_{\rm flat}(\tau)\left(1-{(d-2)\over
24}\tau^2+\dots\right)\, ,\\
A_{\rm dS}(\tau)&=&A_{\rm flat}(\tau)\left(1-{(d-2)\over
12}\tau^2+\dots \right) \, ,\ee where $A_{\rm
flat}(\tau)=Vol(S_{d-2})({\tau\over 2})^{d-2}$, so that we have in
terms of the curvature 
\be A(\tau)=A_{\rm flat}\left(1-{1\over
24(d-1)}R\tau^2+{(d-4)\over 24(d-1)}R_{00}\tau^2+\dots \right)\, .
\ee Notice that the directional component in the expansion
disappears only in dimension $d=4$.

\bigskip

\noindent{\it The iso-perimetric ratios.} The iso-perimetric ratio
can be now calculated, \be &&V_{(d-1)}/A^{d-1\over d-2}\,
(A^{d-1\over
d-2}/V_{(d-1)})_{\rm flat}\nonumber \\
 &&=1+{1\over
8(d-2)(d+1)}R\tau^2+{1\over
4(d-2)(d+1)}R_{00}\tau^2+\dots \nonumber \\
&&= 1+{1\over 4(d-2)(d+1)}(R_{\mu\nu}-{1\over 2}g_{\mu\nu}R)T^\mu
T^\nu+\dots
\nonumber \\
&&=1+{2\pi G\over (d-2)(d+1)}T_{\mu\nu}T^\mu T^\nu+\dots \ee This
extends the results obtained earlier in this paper for $d=4$ to
arbitrary dimension and demonstrates that the iso-perimetric ratio
always involves the energy density in the center of the causal
diamond. It may be that this dependence is a consequence of the
Raychaudhuri equation. It would be interesting to investigate this
possibility  further \footnote{the first author thanks Thibault
Damour for suggesting that this  result on the isoperimetric ratio
might be dimension independent}.

The other iso-perimetric ratio involves the d-volume of the causal
diamond and the maximal (d-1)-volume  \be V_d/V^{d\over
d-1}_{(d-1)}\, (V_{(d-1)}^{d\over d-1}/V_d)_{\rm
flat}=\left(1+{d\over 8(d+1)(d-1)(d+2)}R\tau^2+\dots \right)\, .
\ee This indicates that the directional component vanishes in this
ratio universally in any dimension $d$.

\section{Conclusion}
In this paper we have provided universal formulae for  the volume
and other geometric quantities of small causal diamonds in terms
of the local values of the Ricci tensor and its derivatives
correcting the 't Hooft's flat space values. In all  spacetime
dimensions, our corrections  are valid to quadratic order in the
duration $\tau$ of the causal diamond. In two spacetime dimensions
we are able to work to quartic order in $\tau$. Going to fourth
order and beyond in higher than two spacetime dimensions  appears
to be rather more challenging because of the number of allowed
terms that may contribute.

In general the geometry of causal diamonds turns out to be related
to the distribution of energy and momentum in a non-obvious and in
general directional fashion. However some general trends may be
observed. For instance, one striking result was the behaviour of
one of the  isoperimetric ratios which depends on just the local
energy density in all spacetime dimensions. In the case of the the
other isoperimetric ratio there is no directional behaviour in all
spacetime dimensions. It is hoped that these results will
contribute to a more quantitative understanding of holography and
of  probabilities in eternal inflationary models.

\section{Acknowledgements}
The first  author would like to thank Jan Myrheim and    Raphael
Sorkin for helpful discussions at a very early stage of this work.
Part of it  was carried out at the IH\'ES. Both authors would like
to thank Thibault Damour and the director Jean Pierre Bourguignon,
for their hospitality during our stay at IH\'ES.  Research of the
second author is supported in part by DFG.


\begin{thebibliography}{99}

\bibitem{Sorkin}
 R.~D.~Sorkin,
Causal sets: Discrete gravity,
 [arXiv:gr-qc/0309009].

\bibitem{Bousso}
  R.~Bousso, R.~Harnik, G.~D.~Kribs and G.~Perez,
Predicting the cosmological constant from the causal entropic
principle, [arXiv:hep-th/0702115].

\bibitem{'tHooft} G. 't Hooft, Quantum gravity:
a fundamental problem and some radical ideas", in {\it Recent
Developments in Gravitation} (Proceedings of the 1978 Cargese
Summer Institute) edited by M. Levy and S. Deser (Plenum, 1979).


\bibitem{HawkingKingMcCarthy}
 S W Hawking, A King and P McCarthy, A new topology for curved space-time which
incorporates the causal, differential and conformal structures
{\it J Math Phys} {\bf 17} (1976) 174.

\bibitem{Malament} D. Malament, The class of continuous timelike paths
determines the topology of spacetime { \it J Math Phys} {\bf 18}
(1977) 1399.
\bibitem{Myrheim} J Myrheim,  Statistical Geometry, CERN preprint
TH.25 38-CERN (1978).






\end{thebibliography}
\end{document}